\documentclass[seceq]{ptptex}

\newcommand{\EQ}{\begin{equation}}
\newcommand{\EN}{\end{equation}}
\newcommand{\bea}{\begin{eqnarray}}
\newcommand{\ena}{\end{eqnarray}}
\newcommand{\bdis}{\begin{displaymath}}
\newcommand{\edis}{\end{displaymath}}

\renewcommand{\a}{\alpha}
\renewcommand{\b}{\beta}

\renewcommand{\t}{\tau}

\newcommand{\nn}{\nonumber \\}

\title{
Derivation of Superconformal Anomaly without Ghosts \\
 in ${\cal N}$ = 1 SYM$_4$
}

\author{
Naohito\ Nakazawa
}

\inst{
 Theoretical Physics Laboratory, RIKEN, \\
 Wako 351-0198, Japan
}

\abst{
The anomalous Ward-Takahashi identity for the superconformal symmetry in the four-dimensional ${\cal N}=1$ supersymmetric Yang-Mills theory is studied in terms of the stochastic quantization method (SQM). By applying the background field method to the SQM approach, we derive the superconformal anomaly in the one-loop approximation and show that the supersymmetric stochastic gauge fixing term does not contribute to the anomaly. 
}

\begin{document}

\maketitle

\section{Introduction}

It is well known that the superconformal symmetry is anomalous in ${\cal N}=1$ supersymmetric Yang-Mills theory (SYM$_4$).\cite{AGS} In the conventional (path-integral) BRST invariant formulation\cite{FP} in terms of the superfield,\cite{FZ} the expectation value of the supertrace of the superconformal current, i.e., the superconformal anomaly, is independent of the gauge fixing procedure.\cite{PiSi,Mar,GW-GMZ,SV,Mehta,HOOS} This is ensured manifestly by the BRST exact form of the contribution from the gauge fixing and ghosts in the Ward-Takahashi (W-T) identity for the superconformal symmetry.\cite{HOOS} In this paper, we study this anomalous superconformal W-T identity in the context of the SQM approach. The W-T identity is derived from the superfield Langevin equation with the supersymmetric stochastic gauge fixing procedure, which is formally equivalent to the conventional Faddeev-Popov prescription.\cite{Nakazawa1-2}
 We show that there is also no contribution to the superconformal anomaly from the stochastic gauge fixing term, as a result of an analogous self-cancellation mechanism that corresponds to the BRST exactness in the conventional BRST invariant formulation.
We first derive the anomalous superconformal W-T identity by applying SQM to SYM$_4$ in the superfield formalism. Then, we demonstrate that the SQM approach is equivalent to the conventional BRST invariant approach.

\section{Anomalous Ward-Takahashi identity in Ito calculus}

In the SQM approach, the superfield Langevin equation for SYM$_4$ is given by\cite{Nakazawa1-2}
\bea
\label{eq:Langevin-noise-eq1} 
\displaystyle{\frac{1}{2g}}
\left( \Delta e^{2g{\hat V}}
\right) e^{-2g{\hat V}}  
&+& \beta\Delta \t \displaystyle{\frac{i}{2}} \left(
{\hat {\overline \Phi}} 
- e^{2g L_{\hat V}} {\hat \Phi} \right)   \nn         
& = &
- \beta\Delta\t \displaystyle{\frac{1}{2g}}  \left(
e^{2g L_{\hat V}} {\hat {\cal D}}^\a {\hat W}_\a + {\hat {\overline{\cal D}}}_{\dot \a}{\hat {\overline W}}^{\dot \a}  \right)
+ \Delta w                  \ , \nn
\langle \Delta w^a ( \t, z ) \Delta w^b ( \t, z' )
\rangle_{\Delta w_\t}   
& = &
  \beta\Delta \t 2 \delta^{ab} \delta^8 ( z - z' )        \ .
\ena
Here $L_{\hat V} X \equiv [{\hat V},\ X]$ and the stochastic gauge fixing functions are defined by
$
{\hat \Phi}
 = i {\xi\over 4}
 {\overline D}^2 D^2 {\hat V}
$
and
$
{\hat {\overline \Phi}}
 = - i {\xi\over 4}
 D^2 {\overline D}^2 {\hat V}  
$.
The parameter $\beta$ is introduced as the scale of the stochastic time $\t$.
Discretized notation is used for the time evolution to allow a clear understanding of the essence of It${\bar {\rm o}}$ calculus. 
The notation $\langle ... \rangle_{\Delta w_\t}$ denotes the expectation value evaluated by means of the noise correlation at $\t$. 
We use the convention introduced in Ref.\citen{Nakazawa3}\footnote{In particular, we use the abbreviated notation 
${\hat {\cal D}}^\a {\hat W}_\a
\equiv \{ {\hat {\cal D}}^\a,\  {\hat W}_\a \} $
for
$[ t^a,\ t^b ] = i f^{abc} t^c$, with
${\rm tr}(t^a t^b) = {\rm k}\delta^{ab}$ and
${\rm Tr} \equiv {1\over {\rm k}}{\rm tr}$,
in the sense that 
$
\{ {\hat {\cal D}}^\a,\  {\hat W}_\a \}
= ({\hat {\cal D}}^\a)^{ab} {\hat W}_\a^b t^a
$
in the adjoint representation normalized as
${\rm tr} (T^a T^b) = C_2 (G) \delta^{ab}$
with
$(T^a)_{bc} \equiv i f^{bac} $. 
} in which ${\hat {\cal O}}$ denotes the original vector field and functions of it, which should be distinguished from the quantum fluctuations around the background field.

The superconformal current is derived using a variational principle. We consider the variation of the action 
$
{\hat S}
 =  
- \int\!\! d^8z {1\over 4 g^2} {\rm Tr}    (
{\hat W}^\alpha {\hat W}_\alpha \delta^2 ( {\bar \theta} ) + {\hat {\overline W}}_{\dot \alpha}{\hat {\overline W}}^{\dot \alpha} \delta^2 ( \theta ) )
$ under the $local$ superconformal transformation ${\hat \delta}_{\rm sc}$,\cite{Shizuya-Lang}
\bea
\label{eq:superconformal-transf-eq1}
{\hat \delta}_{\rm sc} e^{2g{\hat V}}
=
-2
\Big(
e^{2g{\hat V}} \Omega^\a {\hat W}_\a
+
{\overline \Omega}_{\dot \a}{\hat {\overline W}}^{\dot \a} e^{2g{\hat V}}
\Big)       \ ,
\ena
where $\Omega_\a (z)$ and ${\overline \Omega}_{\dot \a} (z)$ are unconstrained superfields.
From (\ref{eq:Langevin-noise-eq1}) and the variation of the action under (\ref{eq:superconformal-transf-eq1}), we obtain the $stochastic$ W-T identity
\bea
\label{eq:stochastic-W-T-eq1}
{ } & {} &  
 \beta \Delta\t \int\!\!\! d^8z
( D^\a {\overline \Omega}^{\dot \a} - {\overline D}^{\dot \a} \Omega^\a ) \langle {\hat R}_{\a {\dot \a}} \rangle^\t             \nn
& {} & \quad =
 \displaystyle{\frac{1}{g^2}} \int\!\!\! d^8z \Big\langle
{\overline \Omega}^{\dot \a} {\rm Tr} \Big[ {\hat {\overline W}}_{\dot \a}  \Big\{
\left( \Delta e^{2g{\hat V}}
\right) e^{-2g{\hat V}}  
+ i \beta g \Delta\t \left(
{\hat {\overline \Phi}} - e^{2g L_{\hat V}} {\hat \Phi} \right)   
 \Big\}
   \Big]                   \nn
& {} & \qquad \qquad \quad
- \Omega^\a
{\rm Tr} \Big[  \Big\{
 e^{-2g{\hat V}} \left( \Delta e^{2g{\hat V}}
\right)
+ i \beta g \Delta\t \left(
e^{-2g L_{\hat V}} {\hat {\overline \Phi}} - {\hat \Phi} \right)   
\Big\} {\hat W}_\a  \Big]   \Big\rangle^\t     \ ,
\ena
for the superconformal current,
$
{\hat R}_{\a {\dot \a}}
\equiv - {2\over g^2}
{\rm Tr} {\hat W}_\a e^{-2g{\hat V}} {\hat {\overline W}}_{\dot \a} e^{2g {\hat V}}       \
$.
Here we have used the reality condition,
$
e^{2g L_{\hat V}} {\hat {\cal D}}^\a {\hat W}_\a = {\hat {\overline{\cal D}}}_{\dot \a}{\hat {\overline W}}^{\dot \a} \
$. 
The expectation value 
$\langle ... \rangle^\t$ is understood to be taken by accounting for the correlations of all the noise superfields $\{ \Delta w |  \Delta w( \t' ),\ \t' \leq \t \}$. 
We have also used the fact that the noise variable is uncorrelated with the equal stochastic-time dynamical variables in It${\bar {\rm o}}$ calculus, i.e.
$
\langle {\hat V} ( \t ) \Delta w ( \t )
\rangle^\t  = 0
$. 

We remind that the stochastic gauge fixing term is introduced as a variational term under the local gauge transformation, 
\bea
{\hat \delta}_{\rm lg} {\hat V} 
= - \displaystyle{\frac{i}{2}} {\hat {\cal L}} \Bigl( 
{\hat {\overline \Phi}} - e^{2gL_{\hat V}}{\hat \Phi}  
\Bigr)        \ ,   
\ena
where ${\hat {\cal L}} \equiv 2gL_{\hat V} (e^{2g L_{\hat V}}-1)^{-1}$. 
Therefore, the W-T identity (\ref{eq:stochastic-W-T-eq1}) can be reexpressed as 
\bea
\label{eq:stochastic-W-T-eq2}
{ } & {} &  
 \beta \Delta\t \int\!\!\! d^8z
( D^\a {\overline \Omega}^{\dot \a} - {\overline D}^{\dot \a} \Omega^\a ) \langle {\hat R}_{\a {\dot \a}} \rangle^\t             \nn
& {} & \quad =
 2 \int\!\!\! d^8z \Big\langle
{\rm Tr} \Big[ \Bigl( {\hat {\cal L}}^{\dagger\ -1} {\hat {\cal L}}^{-1} {\hat \delta}_{\rm sc} {\hat V} \Bigr) 
\Bigl( 
\Delta {\hat V} - \beta \Delta\t {\hat \delta}_{\rm lg} {\hat V} 
\Bigr) 
   \Big]         \Big\rangle^\t     \ .
\ena
We note that ${\hat \Phi}$ and ${\hat {\overline \Phi}}$ are the specified gauge fixing functions in SQM. 
In the context of It${\bar {\rm o}}$ stochastic calculus, the origin of the anomalies may be traced to the contact interaction of the derivative terms with respect to the stochastic time.\cite{NOTY} In (\ref{eq:stochastic-W-T-eq1}) and (\ref{eq:stochastic-W-T-eq2}), we show that the anomaly comes from the contact term proportional to $\langle {\hat \delta}_{\rm sc}{\hat V} \Delta {\hat V} \rangle^\t$, while there is no contribution from the stochastic gauge fixing term, $\langle {\hat \delta}_{\rm sc}{\hat V} {\hat \delta}_{\rm lg} {\hat V} \rangle^\t$.

\section{Evaluation of the superconformal anomaly \\
in the background field method}

In order to evaluate the r.h.s. of (\ref{eq:stochastic-W-T-eq1}) in the one-loop approximation, we apply the background field method (BFM)\cite{Abbott,GSR} to the SQM approach.\cite{Nakazawa3} In this method, the original vector superfield
${\hat V}$ is split into its background, ${\bf \Omega}$ and ${\bf \Omega}^\dagger$, and the quantum fluctuations, $V$, as
$e^{2g{\hat V}} = e^{g{\bf \Omega}}e^{2gV}e^{g{\bf \Omega}^\dagger}$.\cite{GSR} The superconformal transformation (\ref{eq:superconformal-transf-eq1}) is also split into its background part, $\delta^{(B)}_{\rm sc}$, and the quantum part, $\delta^{(Q)}_{\rm sc}$, as
${\hat \delta}_{\rm sc} = \delta^{(B)}_{\rm sc} + \delta^{(Q)}_{\rm sc}$: 
\bea
\label{eq:B-Q-split-superconformal-transf-eq1}
\delta_{\rm sc}^{(B)} e^{g{\bf \Omega}}
& = & - 2 e^{g{\bf \Omega}} {\overline \Omega}_{\dot \a} {\overline W}^{{\dot \a}(0)}  
 \ , \quad
\delta_{\rm sc}^{(B)} e^{g{\bf \Omega}^\dagger}
=  - 2  \Omega^\a W_\a^{(0)} e^{g{\bf \Omega}^\dagger}  \ , \nn
\delta^{(Q)}_{\rm sc} e^{2gV}
& = &
-2
\Big(
e^{2gV} \Omega^\a ( W_\a - W_\a^{(0)} )
+
{\overline \Omega}_{\dot \a} ( {\overline W}^{\dot \a} - {\overline W}^{{\dot \a}(0)} ) e^{2gV}
\Big)              \ .  
\ena
Here
${\hat W}_\a = e^{-gL_{{\bf \Omega}^\dagger}} W_\a$,
${\hat {\overline W}}_{\dot \a} = e^{gL_{\bf \Omega}} {\overline W}_{\dot \a}$,  
$W^{(0)}_\a = W_\a |_{V=0} $ and
${\overline W}^{(0)}_{\dot \a} = {\overline W}_{\dot \a} |_{V=0}$.
We expand the non-linear transformation $\delta^{(Q)}_{\rm sc}$ with respect to the quantum fluctuation $V$ as
\bea
\delta^{(Q)}_{\rm sc} V
&=&
\displaystyle{\frac{1}{4}}\Omega^\a ( {\overline {\cal D}}^2{\cal D}_\a V )
+ \displaystyle{\frac{1}{4}}{\overline \Omega}_{\dot \a}( {\cal D}^2{\overline {\cal D}}^{\dot \a} V ) + {\cal O}(V^2) + ...  \ , \nn
&\equiv&
\delta^{(Q1)}_{\rm sc} V + \delta^{(Q2)}_{\rm sc} V + ... \ ,
\ena
where 
${\cal D}_\a \equiv e^{-g{\bf \Omega}}D_\a e^{g{\bf \Omega}}$ and 
${\overline {\cal D}}_{\dot \a} \equiv e^{g{\bf \Omega}^\dagger}{\overline D}_{\dot \a} e^{-g{\bf \Omega}^\dagger}$ 
are the  background covariant spinor derivatives.
The superconformal current and the action are also expanded as 
$
{\hat R}_{\a{\dot \a}}
\equiv
R_{\a{\dot \a}}^{(0)} + R_{\a{\dot \a}}^{(1)} + R_{\a{\dot \a}}^{(2)} + ... \
$ 
and
$
{\hat S} 
\equiv 
S^{(0)} + S^{(1)} + S^{(2)} + ... \
$,
respectively.
The supertrace of the current $R_{\a{\dot \a}}^{(2)}$, which is relevant in the one-loop approximation, is given by  
\bea
\label{eq:conservation-R2-eq1}
\Big( \delta^{(B)}_{\rm sc} + \delta^{(Q1)}_{\rm sc} \Big) S^{(2)}
+ \delta^{(Q2)}_{\rm sc} S^{(1)}
 =
- \displaystyle{\frac{1}{2}} \int\!\!\! d^8z
( D^\a {\overline \Omega}^{\dot \a} - {\overline D}^{\dot \a} \Omega^\a ) R^{(2)}_{\a {\dot \a}}              \ .
\ena

In BFM applied to SYM$_4$ in the SQM approach,\cite{Nakazawa3} the Langevin equation (\ref{eq:Langevin-noise-eq1}) is reduced to
\bea
\label{eq:Langevin-BFM-eq1}
\Delta V
& = &
- \displaystyle{\frac{\beta}{2g}} {\cal L}^\dagger \Big(
\displaystyle{\frac{1}{\beta}} ( \Delta e^{g{\bf \Omega}^\dagger} ) e^{-g{\bf \Omega}^\dagger }  -ig \Delta\t \phi + \Delta\t \nabla^\a W_\a    \Big)
           \nn
& {} & \quad
- \displaystyle{\frac{\beta}{2g}} {\cal L} \Big(
\displaystyle{\frac{1}{\beta}} e^{- g{\bf \Omega}} (\Delta e^{g{\bf \Omega}} )
+ ig \Delta\t {\overline \phi}
+ \Delta\t {\overline \nabla}_{\dot \a}{\overline W}^{\dot \a}  \Big)
+ {\cal L} e^{-gL_{\bf \Omega}}\Delta w                  \ ,
\ena
where 
${\cal L} \equiv 2gL_V (e^{2g L_V}-1)^{-1}$,
$\nabla_\a \equiv e^{-2gV}{\cal D}_\a e^{2gV}$
and
${\overline \nabla}_{\dot \a} \equiv e^{2gV}{\overline {\cal D}}_{\dot \a}e^{-2gV}$.
The background local gauge invariant stochastic gauge fixing functions are given by
$
\phi
 = i {\xi\over 4}
 {\overline {\cal D}}^2 {\cal D}^2 V   
$
and
$
{\overline \phi}
 = - i {\xi\over 4}
 {\cal D}^2 {\overline {\cal D}}^2 V   
$. Since the expectation value of the current $\langle R_{\a{\dot \a}}^{(2)} \rangle^\t$ includes
$\delta^{(B)}_{\rm sc}S^{(2)}$, the quadratic terms in (\ref{eq:Langevin-BFM-eq1}) with respect to $V$ are also relevant in the one-loop approximation. To eliminate the terms that depend on only the background fields in (\ref{eq:Langevin-BFM-eq1}), we have imposed the $stochastic$ equation of motion 
\bea
\label{eq:stochastic-eq-motion-eq1}
\displaystyle{\frac{1}{\beta}} \left( \displaystyle{\frac{d}{d\t}} e^{g{\bf \Omega}^\dagger} \right) e^{-g{\bf \Omega}^\dagger } + \displaystyle{\frac{1}{\beta}} e^{- g{\bf \Omega}} \left( \displaystyle{\frac{d}{d\t}} e^{g{\bf \Omega}} \right)
+ {\cal D}^\a W_\a^{(0)} + {\overline {\cal D}}_{\dot \a}{\overline W}^{{\dot \a}(0)} = 0         \ .
\ena
The background fields, ${\bf \Omega}$ and ${\bf \Omega}^\dagger$, must be independent of the stochastic time in the equilibrium limit. This means that the classical equations of motion, 
$
{\cal D}^\a W_\a^{(0)} = {\overline {\cal D}}_{\dot \a}{\overline W}^{{\dot \a}(0)} = 0  
$, 
hold for the background fields in the equilibrium limit. 
This requirement is sufficient for the equivalence of the SQM approach and the conventional one. For practical reasons, we simply assume that the background field is independent of the stochastic time, i.e., 
${\dot {\bf \Omega}} = {\dot {\bf \Omega}}^\dagger = 0$, 
and impose the classical equations of motion for the background fields 
$at\ finite\ stochastic\ time$. This assumption and (\ref{eq:stochastic-eq-motion-eq1})  
hold only in the tree approximation. In the one-loop approximation, as we show in the section 4, we impose the effective $stochastic$ equation of motion as a Schwinger-Dyson (S-D) equation derived from the expectation value of (\ref{eq:Langevin-noise-eq1}). Here we note that the use of the classical background equations of motion in evaluating (\ref{eq:stochastic-W-T-eq1}) is consistent with the one-loop approximation. In the one-loop approximation, the W-T identity (\ref{eq:stochastic-W-T-eq1}) is reduced to
\bea
\label{eq:BFM-stochastic-W-T-eq1}
{ } & {} &  
- \displaystyle{\frac{1}{2}} \beta \Delta\t \int\!\!\! d^8z
( D^\a {\overline \Omega}^{\dot \a} - {\overline D}^{\dot \a} \Omega^\a ) \langle  R^{(2)}_{\a {\dot \a}} \rangle^\t             \nn
& {} &
\qquad \qquad =
 \int\!\!\! d^8z \Big\{
- \displaystyle{\frac{1}{g}}{\overline \Omega}^{\dot \a} \Big\langle {\rm Tr} ( {\overline W}^{(1)}_{\dot \a} \Delta V ) \Big\rangle^\t  
+ \displaystyle{\frac{1}{g}} \Omega^\a  
\Big\langle
{\rm Tr} ( \Delta V W^{(1)}_\a ) \Big\rangle^\t     \nn
& {} & \qquad \qquad \qquad
+ i \beta \Delta\t   {\overline \Omega}^{\dot \a} \Big\langle {\rm Tr} \Big(
{\overline W}_{\dot \a}^{(0)} [V,\ \phi ]
- \displaystyle{\frac{1}{2g}} {\overline W}_{\dot \a}^{(1)} ( {\overline \phi} - \phi )        \Big)  \Big\rangle^\t       \  \nn
& {} & \qquad \qquad \qquad
- i \beta \Delta\t   \Omega^\a  \Big\langle {\rm Tr} \Big(
 [V,\ {\overline \phi} ] W_\a^{(0)}
- \displaystyle{\frac{1}{2g}} ( {\overline \phi} - \phi ) W_\a^{(1)}   
    \Big) \Big\rangle^\t  \Big\}         \ .
\ena
To derive this expression, we have used the fact that the terms that remain non-trivial in (\ref{eq:BFM-stochastic-W-T-eq1}) must produce at least four spinor covariant derivatives, 
${\cal D}_\a {\cal D}_\b {\overline {\cal D}}_{\dot \a} {\overline {\cal D}}_{\dot \b}$, after the regularization procedure.

We first evaluate the term proportional to $\langle V\Delta V \rangle^\t$ in (\ref{eq:BFM-stochastic-W-T-eq1}) by introducing the Gaussian cut-off regularization for the noise correlation as
\bea
\label{eq:regularized-noise-eq1}
\langle
\Delta w^a (\t, z) \Delta w^b (\t, z')
\rangle_{\Delta w_\t} |_{\rm reg}
=
2\beta \Delta\t \langle a, z| e^{{\hat \square}_\xi / \Lambda^2} | b,z' \rangle , \nn
\square_\xi
\equiv
{\cal D}^m {\cal D}_m - W^{\a(0)} {\cal D}_\a  
+ {\overline W}^{(0)}_{\dot \a} {\overline {\cal D}}^{\dot \a}
+ \displaystyle{\frac{\xi-1}{16}} ( {\overline {\cal D}}^2{\cal D}^2 + {\cal D}^2{\overline {\cal D}}^2 )       \ .
\ena
Here we have introduced the bra-ket notation 
$
\langle a,z | b,z' \rangle = \delta^{ab}\delta^8(z-z')
$. 
The operator ${\hat \square}_\xi$
is an abstract operator defined by
$\langle a, z | {\hat \square}_\xi | b, z' \rangle
= ( \square_\xi )^{ac}_z \langle c, z | b, z' \rangle $ in the adjoint representation.  
The differential operator $\square_{\xi}$ has been determined from the kinetic term in (\ref{eq:Langevin-BFM-eq1}). 
The symbol \lq\lq $\small\wedge$ \rq\rq in the bra-ket notation simply represents the abstract operator of the differential operator. 
From (\ref{eq:Langevin-BFM-eq1}) and (\ref{eq:regularized-noise-eq1}),
we obtain the regularized superpropagator of the vector superfield: 
\bea
\label{eq:regularized-vector-superpropagator-eq1}
& {} &
\langle V^a (\t, z) V^b (\t', z') \rangle^{{\rm max}(\t, \t')}_{\rm reg}             \nn
& {} & \qquad \qquad
=   
\langle a,z | e^{ - 2\beta \t {\hat \square}_\xi}   V(0)V(0) e^{ - 2\beta \t' {\hat \square}_\xi}  
 | b,z' \rangle        \nn
& {} & \qquad \qquad \quad
+ \displaystyle{\frac{1}{2}} \langle a,z |
 (- {\hat \square}_\xi)^{-1} \Big( e^{ -2\beta (\t+\t') {\hat \square}_\xi}  
- e^{ -2\beta | \t-\t'| {\hat \square}_\xi} \
\Big) e^{{\hat \square}_\xi / \Lambda^2}
 | b,z' \rangle       \ .
\ena
Here, $V(0)$ denotes the initial value of the vector superfield to solve the Langevin equation.
In the equilibrium limit, the dependence on the initial conditions vanishes due to the stochastic gauge fixing procedure,\footnote{To reproduce the normal sign factor in the component Langevin equations and the component noise correlations, we redefine the scaling parameter $\beta$ as $\beta = - {1\over 2\kappa}$,\cite{Nakazawa3} which also ensures the damping of the initial condition dependence in (\ref{eq:regularized-vector-superpropagator-eq1}) in the equilibrium limit.}  
\bea
\label{eq:regularized-vector-superpropagator-eq2}
\lim_{\t \rightarrow \infty} \langle V^a (\t, z) V^b (\t, z') \rangle^\t_{\rm reg}     
= 
- \displaystyle{\frac{1}{2}} \langle a,z | (- {\hat \square}_\xi)^{-1} e^{{\hat \square}_\xi / \Lambda^2}   | b,z' \rangle      
 \ .
\ena
In the Gaussian cut-off regularization procedure, we take the limit of infinite stochastic time, $\t \rightarrow \infty$, before taking the limit $\Lambda \rightarrow \infty$.  

In the sense of It${\bar {\rm o}}$ stochastic calculus, the essential point with regard to the evaluation of the anomaly is the following. Consider the stationary condition of the stochastic-time evolution in the equilibrium limit, 
$
\Delta \langle V V'\rangle^\t_{\rm reg} = \langle (\Delta V)V' + V(\Delta V') + (\Delta V)(\Delta V') \rangle^\t_{\rm reg} \rightarrow 0$
 at
$\t \rightarrow \infty$.
This implies
\bea
\label{eq:regularized-contact-term}
\langle (\Delta V)V' \rangle^\infty_{\rm reg} |_{z=z'}  
& = &
\langle V(\Delta V') \rangle^\infty_{\rm reg} |_{z=z'}    \  \nn
& = &
- {1\over 2} \langle (\Delta V)(\Delta V') \rangle^\infty_{\rm reg} |_{z=z'}           \  \nn
& = &
- {1\over 2} \langle \Delta w \Delta w' \rangle_{\rm reg} |_{z=z'}    
   + O ( \Delta \t^{3/2} ) \ , 
\ena
where $\langle ... \rangle^\infty \equiv \lim_{\t \rightarrow \infty} \langle ... \rangle^\t$ 
represents the expectation value in the equilibrium limit. 
This relation in the equilibrium limit can be checked directly using the regularized vector superpropagator (\ref{eq:regularized-vector-superpropagator-eq1}). 
From (\ref{eq:regularized-contact-term}), we can evaluate the anomaly terms, the contact terms proportional to $\langle V \Delta V \rangle^\t_{\rm reg}$ in (\ref{eq:BFM-stochastic-W-T-eq1}), as
\bea
\label{eq:anomaly-SQM-eq1}
& {} &
\lim_{\Lambda \rightarrow \infty}
\int\!\!\! d^8z
\Big(
\displaystyle{\frac{1}{4}}\Omega^\a ({\overline {\cal D}}^2{\cal D}_\a )_z^{ab}
+ \displaystyle{\frac{1}{4}}{\overline \Omega}_{\dot \a} ( {\cal D}^2 {\overline {\cal D}}^{\dot \a} )_z^{ab} \Big) \langle b, z | e^{{\hat \square_{(\xi = 1)}}/\Lambda^2 }| a,z' \rangle \Big|_{z=z'}  \nn
& {} & \quad
=
- \int\!\!\! d^8z \displaystyle{\frac{C_2(G)}{2(4\pi)^2}}
\Big\{
\Omega^\a D_\a {\rm Tr}(W^{\b(0)}W_\b^{(0)})
+
{\overline \Omega}_{\dot \a} {\overline D}^{\dot \a}
{\rm Tr}({\overline W}_{\dot \b}^{(0)}
{\overline W}^{{\dot \b}(0)} )
\Big\}             \ ,
\ena
up to a factor of $\beta \Delta \t$.
Here we have performed this standard evaluation of the superconformal anomaly in the Feynman gauge,
$\xi = 1$.
Therefore, provided that the remaining terms in the stochastic W-T identity (\ref{eq:BFM-stochastic-W-T-eq1}) do not contribute to the anomaly,
%
%
we obtain the well-known anomalous superconformal W-T identity for SYM$_4$, 
\bea
\label{eq:SYM-anomalous-W-T-eq1}
\langle {\overline D}^{\dot \a} R_{\a {\dot \a}} \rangle^\infty 
& = & 
- \displaystyle{\frac{C_2(G)}{(4\pi)^2}}
D_\a {\rm Tr}(W^{\b(0)}W_\b^{(0)})        \ , \nn
\langle D^\a R_{\a {\dot \a}} \rangle^\infty            
& = &  
- \displaystyle{\frac{C_2(G)}{(4\pi)^2}}
{\overline D}_{\dot \a}
{\rm Tr}({\overline W}_{\dot \b}^{(0)}
{\overline W}^{{\dot \b}(0)} )       \ ,
\ena
in the one-loop approximation.

We can show that the third and the fourth terms on the r.h.s. of the stochastic W-T identity (\ref{eq:BFM-stochastic-W-T-eq1}) vanish in the equilibrium limit by using the representation
(\ref{eq:regularized-vector-superpropagator-eq2}) as follows: 
\bea
\label{eq:identity-vector-superpropagator-eq1}
& {} &
- \displaystyle{\frac{1}{2g}} \Big\langle
{\rm Tr} \Big\{
 W^{(1)}_\a (z)  ( {\cal D}^2{\overline {\cal D}}^2 + {\overline {\cal D}}^2{\cal D}^2) V(z')
\Big\}
 \Big\rangle^\infty_{\rm reg} \Big|_{z=z'}    \nn
& {} & \qquad
=
\displaystyle{\frac{1}{8}} \Big\langle
{\rm Tr} \Big\{
({\overline {\cal D}}^2 {\cal D}_\a V(z) ) ({\cal D}^2{\overline {\cal D}}^2 V(z') )
\Big\}
 \Big\rangle^\infty_{\rm reg} \Big|_{z=z'}    \nn
& {} & \qquad
= -\displaystyle{\frac{1}{16}} \Big(
[{\overline {\cal D}}_{\dot \a},\ \{ {\overline {\cal D}}^{\dot \a},\
{\cal D}_\a \} ] \Big)_z^{ab}
\langle b,z | \int_0^{\infty}\!\!\! ds e^{s\xi {\hat \square}^{\rm FP} + \xi {\hat \square}^{\rm FP}/ \Lambda^2 }  
{\hat {\overline \nabla}}^2 {\hat \nabla}^2
 | a,z' \rangle \Big|_{z=z'}         \nn
 & {} & \qquad
=
- \Big\langle
{\rm Tr}\Big\{ [W_\a^{(0)},\ V(z)]  
{\cal D}^2{\overline {\cal D}}^2 V(z') \Big\}
\Big\rangle^\infty_{\rm reg} \Big|_{z=z'}     \ .
\ena
Here, the differential operators $\square^{\rm FP}$ and ${\tilde \square}^{\rm FP}$ are defined by
\bea
\label{eq:def-FP-operator-eq1}
\square^{\rm FP}
& \equiv &
{\cal D}^m {\cal D}_m - W^{\a(0)} {\cal D}_\a  
- \displaystyle{\frac{1}{2}} {\cal D}^\a W^{(0)}_\a    \ , \nn
{\tilde \square}^{\rm FP}
& \equiv &
{\cal D}^m {\cal D}_m + {\overline W}^{(0)}_{\dot \a} {\overline {\cal D}}^{\dot \a}
+ \displaystyle{\frac{1}{2}} {\overline {\cal D}}_{\dot \a} {\overline W}^{{\dot \a}(0)}  \ .
\ena
In (\ref{eq:identity-vector-superpropagator-eq1}), the abstract operators
${\hat \square}^{\rm FP}$,
${\hat \nabla}^2$ and
${\hat {\overline \nabla}}^2$ are defined by
$\langle a, z | {\hat \square}^{\rm FP} |b, z' \rangle
\equiv (\square^{\rm FP})^{ac}_z \langle c, z | b, z' \rangle $
,
$
\langle a, z | {\hat \nabla}^2 |b, z' \rangle
\equiv ({\cal D}^2)^{ac}_z \langle c, z | b, z' \rangle
$ and
$\langle a, z | {\hat {\overline \nabla}}^2 |b, z' \rangle
\equiv ({\overline {\cal D}}^2)^{ac}_z \langle c, z | b, z' \rangle$,
respectively. 
The equation (\ref{eq:identity-vector-superpropagator-eq1}) can be shown by the relations
\bea
\label{eq:relation-FP-vector-operator-eq1}
{\cal D}^2 \square_\xi
& = &
\displaystyle{\frac{\xi}{16}} {\cal D}^2 {\overline {\cal D}}^2 {\cal D}^2
= \xi {\cal D}^2 {\tilde \square}^{\rm FP}
= \xi {\tilde \square}^{\rm FP} {\cal D}^2
=  \square_\xi {\cal D}^2    \ ,  \nn
{\overline {\cal D}}^2 \square_\xi      
& = &
\displaystyle{\frac{\xi}{16}} {\overline {\cal D}}^2 {\cal D}^2 {\overline {\cal D}}^2
= \xi {\overline {\cal D}}^2 \square^{\rm FP}
= \xi \square^{\rm FP} {\overline {\cal D}}^2
=  \square_\xi {\overline {\cal D}}^2    \ ,
\ena
which hold under the classical background equations of motion, ${\cal D}^\a W^{(0)}_\a = {\overline {\cal D}}_{\dot \a} {\overline W}^{{\dot \a}(0)} = 0$. For  later convenience, we have introduced the differential operators $\square^{\rm FP}$ and ${\tilde \square}^{\rm FP}$. They are identical to those appearing in the kinetic terms of the F-P ghosts in the conventional BRST formalism in the path-integral method. 
In this section, we have shown that the anomalous superconformal W-T identity (\ref{eq:SYM-anomalous-W-T-eq1}) is recovered in the SQM approach with the Gaussian cut-off regularization. Our analysis is restricted to the W-T identity in the equilibrium limit. This is mainly because of the initial condition dependence of the regularized vector superpropagator given in (\ref{eq:regularized-vector-superpropagator-eq1}). If we ignore this initial condition dependence, then the identity (\ref{eq:identity-vector-superpropagator-eq1}) holds $at$ $finite$ $stochastic$ $time$.
As we  show in the section 5, the identity (\ref{eq:identity-vector-superpropagator-eq1}) is equivalent to the BRST exactness of the contribution from the gauge fixing auxiliary fields and ghost fields in the conventional approach.

\section{Effective equation of motion}

As we have already mentioned, we impose the effective $stochastic$ equations of motion instead of (\ref{eq:stochastic-eq-motion-eq1}) as a S-D equation in the equilibrium limit for the one-loop approximation. We consider the expectation value of the Langevin equation (\ref{eq:Langevin-noise-eq1}). For later convenience, we express it in terms of an arbitrary variation of the action with respect to the vector superfield as
\bea
\label{eq:SQM-pre-effective-eq-motion-eq1}
2g\beta \Delta\t \langle \delta S \rangle^\t_{\rm reg}
& = &
\int\!\!\!d^8z \Big\langle
{\rm Tr}\Big[
\left( \delta e^{2g{\hat V}}
\right) e^{-2g{\hat V}}  \Big\{
- \displaystyle{\frac{1}{2g}}
\left( \Delta e^{2g{\hat V}}
\right) e^{-2g{\hat V}}                   \nn
& {} & \qquad\qquad\qquad\qquad
- \beta\Delta \t \displaystyle{\frac{i}{2}} \left(
{\hat {\overline \Phi}} - e^{2g L_{\hat V}} {\hat \Phi} \right)           
+ \Delta w   \Big\}
\Big] \Big\rangle^\t_{\rm reg}    \ .
\ena
In BFM, the arbitrary variation of the original vector superfield ${\hat V}$ is split into the background variation and the quantum variation,
$\delta = \delta^{(B)} + \delta^{(Q)}$. We consider only the background variation, i.e., $\delta^{(Q)}V = 0$. The background variation of the action is given by 
\bea
\label{eq:Background-variation-action-eq1} 
\delta^{(B)} S 
& = &
\displaystyle{\frac{1}{2g^2}} \int\!\!\! d^8z {\rm Tr}\Big\{
e^{-g{\bf \Omega}} ( \delta^{(B)} e^{g{\bf \Omega}} )
 \{ {\overline \nabla}_{\dot \a},\ {\overline W}^{\dot \a}  \}     \nn
& {} &    \qquad\qquad\qquad
+ ( \delta^{(B)} e^{g{\bf \Omega}^\dagger} ) e^{-g{\bf \Omega}^\dagger} 
\{  \nabla^\a,\ W_\a \}   
\Big\}            \    . 
\ena
Thus, up to ${\cal O}( V^2 )$ for the one-loop approximation, (\ref{eq:SQM-pre-effective-eq-motion-eq1}) reads
\bea
\label{eq:SQM-effective-eq-motion-eq1}
& {} &
\displaystyle{\frac{1}{4g^2}}
\int\!\!\! d^8z {\rm Tr} \Big\{ \Big(
e^{-g{\bf \Omega}} ( \delta^{(B)} e^{g{\bf \Omega}} )
+ ( \delta^{(B)} e^{g{\bf \Omega}^\dagger} ) e^{-g{\bf \Omega}^\dagger}
\Big) \left( {\rm l.h.s.\ of\ (\ref{eq:stochastic-eq-motion-eq1})} \right)  \Big\}       \nn
& {} & \quad
+ \int\!\!\! d^8z {\rm Tr}\Big\{
e^{-g{\bf \Omega}} ( \delta^{(B)} e^{g{\bf \Omega}} )
\Big\langle
\displaystyle{\frac{1}{2g^2}} \{ {\overline \nabla}_{\dot \a},\ {\overline W}^{\dot \a}  \}^{(2)} - \displaystyle{\frac{i}{2}} [V,\ \phi ]
\Big\rangle^\infty_{\rm reg}           \nn
& {} & \qquad \qquad \qquad
+ ( \delta^{(B)} e^{g{\bf \Omega}^\dagger} ) e^{-g{\bf \Omega}^\dagger}
\Big\langle      
\displaystyle{\frac{1}{2g^2}} \{  \nabla^\a,\ W_\a \}^{(2)} - \displaystyle{\frac{i}{2}} [V,\ {\overline \phi} ]
\Big\rangle^\infty_{\rm reg}    
\Big\}    = 0          \ 
\ena
in the equilibrium limit. This is, in fact, equivalent to the effective equations of motion in the conventional approach described by the background variation of the generator of the 1-P-I vertices, provided that
${\dot {\bf \Omega}}= {\dot {\bf \Omega}^\dagger} = 0$ in (\ref{eq:SQM-effective-eq-motion-eq1}).

\section{Equivalence to the BRST invariant path-integral approach}

In the following, we clarify the equivalence of the SQM approach and the conventional BRST invariant approach.\cite{FP} 
We first derive the superconformal W-T identity in the conventional BRST invariant formulation. The generating functional is defined by 
\bea
\label{eq:conventional-BRST-generating-functional-eq0}
& {} &
{\hat Z} = \int\!\!\! d{\hat \mu} e^{- {\hat S}_{\rm tot} + {\hat S}_{\rm ex}} \ , \quad
{\hat S}_{\rm tot} = {\hat S} + {\hat S}_{\rm gf} + {\hat S}_{\rm FP}                   \ , \nn
& {} &
{\hat S}_{\rm gf} + {\hat S}_{\rm FP}
= - i \int\!\!\! d^8z {\hat \delta}_{\rm BRST} {\rm Tr}\Big\{ ({\hat c}'+{\hat {\overline c}}') {\hat V} \Big\}          \nn
& {} & \qquad\qquad\qquad\qquad\qquad 
+  {\rm Tr} \Big\{  
 \int\!\!\! d^6z {\hat B}{\hat f} + \int\!\!\! d^6{\bar z} {\hat {\overline B}}{\hat {\overline f}}  + 2\xi \int\!\!\! d^8z {\hat {\overline f}}{\hat f}  
\Big\}   \ , 
\ena
where
$d{\hat \mu} \equiv {\cal D}{\hat V}{\cal D}{\hat c}{\cal D}{\hat {\overline c}}'{\cal D}{\hat {\overline c}}{\cal D}{\hat c}'{\cal D}{\hat B}{\cal D}{\hat {\overline B}}{\cal D}{\hat f}{\cal D}{\hat {\overline f}}$. ${\hat B}$ and ${\hat {\overline B}}$ are the Nakanishi-Lautrap (N-L) fields, ${\hat f}$ and ${\hat {\overline f}}$ are the gauge averaging functions, and  
 ${\hat S}_{\rm ex}$ denotes the external source terms.
 The nilpotent BRST operator is defined by
\bea
\label{eq:conventional-BRST-operator-eq1} 
{\hat \delta}_{\rm BRST}
& = &
\int \!\! d^8z \Big\{
\left( - \displaystyle{\frac{i}{2}} {\hat L}^a_{\ b}{\hat c}^b
+ \displaystyle{\frac{i}{2}} {\hat {\overline c}}^b {\hat L}_b^{\ a}
 \right) \displaystyle{\frac{\delta}{\delta {\hat V}^a}}
- \displaystyle{\frac{g}{2}} [{\hat c}\times {\hat c}]^a  \displaystyle{\frac{\delta}{\delta {\hat c}^a}}
- \displaystyle{\frac{g}{2}} [{\hat {\overline c}}\times {\hat {\overline c}}]^a  \displaystyle{\frac{\delta}{\delta {\hat {\overline c}}^a}}       \nn
& {} & \qquad \qquad \qquad +
i {\hat {\overline B}}^a \displaystyle{\frac{\delta}{\delta {{\hat {\overline c}}'_a}}}
+ i {\hat B}^a \displaystyle{\frac{\delta}{\delta {\hat c}'_a}}   
   \Big\}             \ ,
\ena
where ${\hat L}^a_{\ b} = {\rm Tr} (t_b \cdot {\hat {\cal L}} t^a )$
and
${\hat L}^{\ a}_b = {\rm Tr}(t_b \cdot {\hat {\cal L}}^\dagger t^a )$. 
The superconformal W-T identity is derived as the Schwinger-Dyson equation 
\bea
\label{eq:conventional-BRSTinvariant-W-T-eq1}
0 
& = & 
{\hat Z}^{-1} \int\!\!\! d{\hat \mu} \int\!\!\! d^8z 
\displaystyle{\frac{\delta}{\delta {\hat V}(z)}} \Bigl( 
{\hat \delta}_{\rm sc} {\hat V}(z) e^{- {\hat S}_{\rm tot}} 
\Bigr)               \ , \nn
& = &  
\Big\langle \int\!\!\! d^8z 
\displaystyle{\frac{\delta}{\delta {\hat V}(z)}}  
{\hat \delta}_{\rm sc} {\hat V}(z) - {\hat \delta}_{\rm sc} {\hat S}_{\rm tot} 
 \Big\rangle^{\rm BRST}            \ . 
\ena
The expectation value $\langle ... \rangle^{\rm BRST} $ is defined by the generating functional ${\hat Z}$. 
All the auxiliary fields and ghosts, though they are chiral or anti-chiral superfields, are assumed to be invariant under the local superconformal transformation. Therefore, the superconformal transformation commutes with the BRST transformation, i.e. 
$
[ {\hat \delta}_{\rm BRST},\  {\hat \delta}_{\rm sc}  ] = 0  \ .
$ 
We thus obtain the anomalous superconformal W-T identity, 
\bea
\label{eq:conventional-BRSTinvariant-W-T-eq2}
& {} &
  - \displaystyle{\frac{1}{2}} \int\!\!\! d^8z
( D^\a {\overline \Omega}^{\dot \a} - {\overline D}^{\dot \a} \Omega^\a ) \langle {\hat R}_{\a {\dot \a}} \rangle^{\rm BRST}              \nn
& {} & \quad
=  
 \int\!\!\! d^8z \Big\langle
\displaystyle{\frac{\delta}{\delta {\hat V}(z)}}  
{\hat \delta}_{\rm sc} {\hat V}(z) 
 \Big\rangle^{\rm BRST}           
+ i \!\!\int\!\!\! d^8z \Big\langle
{\hat \delta}_{\rm BRST} \Big\{
 {\rm Tr}(({\hat c}' + {\hat {\overline c}}' ) {\hat \delta}_{\rm sc} {\hat V}) \Big\} \Big\rangle^{\rm BRST}            \ .
\ena
For the explicit evaluation of the r.h.s., we need to specify a regularization procedure. Under an appropriate regularization, by comparing this expression for the anomalous superconformal W-T identity to (\ref{eq:stochastic-W-T-eq2}), we expect the following identities: 
\bea
\label{eq:stochastic-BRST-correspondence-eq1}
 - \displaystyle{\frac{1}{\beta\Delta\t}}  \Big\langle
{\rm Tr} \Big\{ \Bigl( {\hat {\cal L}}^{\dagger\ -1} {\hat {\cal L}}^{-1} {\hat \delta}_{\rm sc} {\hat V} \Bigr) 
\Bigl( 
\Delta {\hat V} 
\Bigr) 
   \Big\}         \Big\rangle^\t_{\rm reg} 
& = & 
 \Big\langle
\displaystyle{\frac{\delta}{\delta {\hat V}(z)}}  
{\hat \delta}_{\rm sc} {\hat V}(z) 
 \Big\rangle^{\rm BRST}_{\rm reg}                     \ , \\ 
\label{eq:stochastic-BRST-correspondence-eq2} 
  \Big\langle
{\rm Tr} \Big\{ \Bigl( {\hat {\cal L}}^{\dagger\ -1} {\hat {\cal L}}^{-1} {\hat \delta}_{\rm sc} {\hat V} \Bigr) 
\Bigl( 
 {\hat \delta}_{\rm lg} {\hat V} 
\Bigr) 
   \Big\}         \Big\rangle^\t_{\rm reg}  
& = & 
 i \Big\langle
{\hat \delta}_{\rm BRST} \Big\{
 {\rm Tr}(({\hat c}' + {\hat {\overline c}}' ) {\hat \delta}_{\rm sc} {\hat V}) \Big\} \Big\rangle^{\rm BRST}_{\rm reg}            \ .
\ena   
 In particular, the r.h.s. of (\ref{eq:stochastic-BRST-correspondence-eq2}) is the expectation value of a BRST exact form. Therefore, it must vanish with a BRST invariant vacuum. Accordingly, the l.h.s of (\ref{eq:stochastic-BRST-correspondence-eq2}) must vanish to all orders in the perturbative expansion, i.e., the stochastic gauge fixing term does not contribute to the superconformal W-T identity.  The equivalence of the SQM approach and the conventional BRST invariant one is formally established with a particular choice of the gauge fixing functions ${\hat \Phi}$ and ${\hat {\overline \Phi}}$.\cite{Nakazawa1-2} Here, with the gauge fixing functions given in (\ref{eq:Langevin-noise-eq1}), which are more convenient for actual calculations, we directly check the equivalence of the two superconformal W-T identities, (\ref{eq:stochastic-W-T-eq2}) and (\ref{eq:conventional-BRSTinvariant-W-T-eq2}), in the one-loop approximation on the basis of the background field method (BFM).  

The conventional BRST invariant BFM is defined by introducing the background chiral and background anti-chiral superfields. They are defined in terms of their original superfields as 
$\varphi \equiv e^{gL_{{\bf \Omega}^\dagger}} {\hat \varphi}$
and
${\bar \varphi} \equiv e^{-gL_{\bf \Omega}} {\hat {\bar \varphi}}$. This procedure causes a nontrivial contribution of the gauge averaging functions in  (\ref{eq:conventional-BRST-generating-functional-eq0}), and therefore we need to add the Nielsen-Kallosh (N-K) ghost term. The background chiral (anti-chiral) superfields satisfy the background chiral (anti-chiral) condition, 
${\bar {\cal D}}\varphi = ( {\cal D}{\bar \varphi} ) = 0$, 
where
$\varphi = ( c, c', B, f, b )$
and
${\bar \varphi} = ({\bar c}', {\bar c}, {\bar B}, {\bar f}, {\bar b} )$. 
Here, $b$ and ${\overline b}$ are the Nielsen-Kallosh (N-K) ghosts. 
If we do not integrate out the  N-L fields, then the gauge averaging functions and N-K ghosts cancel each other in (\ref{eq:conventional-BRST-generating-functional-eq0}), and the generating functional of SYM$_4$ is reduced to\cite{HOOS}
\bea
\label{eq:conventional-BRST-generating-functional-eq1}
& {} &
Z = \int\!\!\! d\mu e^{- S_{\rm tot} + S_{\rm ex}} \ , \quad
S_{\rm tot} = S + S_{\rm gf} + S_{\rm FP} + S_{\rm NL}   \ , \nn
& {} &
S_{\rm gf} + S_{\rm FP}
= - i \int\!\!\! d^8z \delta_{\rm BRST} {\rm Tr}\Big\{ (c'+{\overline c}') V \Big\}        \ ,   \nn
& {} &
S_{\rm NL}
= \displaystyle{\frac{1}{2\xi}} \int\!\!\! d^8z {\rm Tr} {\overline B} (- {\tilde \square}^{\rm FP})^{-1} B    \ ,
\ena
where
$d\mu \equiv {\cal D}V{\cal D}c{\cal D}{\overline c}'{\cal D}{\overline c}{\cal D}c'{\cal D}B{\cal D}{\overline B}$. 
The BRST operator $\delta_{\rm BRST}$ is given by ${\hat \delta}_{\rm BRST}$ in (\ref{eq:conventional-BRST-generating-functional-eq0}), substituting the original superfields with their quantum fluctuations.   
The differential operator ${\tilde \square}^{\rm FP}$ has already been defined in (\ref{eq:def-FP-operator-eq1}). 

If we first integrate out the N-L fields and then integrate out the gauge averaging functions, the N-K ghost term remains and 
(\ref{eq:conventional-BRST-generating-functional-eq0}) is expressed in the familiar form\cite{GSR}
\bea
\label{eq:conventional-BRST-generating-functional-eq2}
& {} &
{\tilde Z} = \int\!\!\! d{\tilde \mu} e^{- {\tilde S}_{\rm tot} + {\tilde S}_{\rm ex}} \ , \quad
{\tilde S}_{\rm tot} = S + {\tilde S}_{\rm gf} + S_{\rm FP} + S_{\rm NK}   \ , \nn
& {} &
{\tilde S}_{\rm gf}
= \displaystyle{\frac{\xi}{8}} \int\!\!\! d^8z {\rm Tr}\Big\{ ({\overline {\cal D}}^2 V) ({\cal D}^2 V) \Big\}        \ ,   \nn
& {} &
S_{\rm NK}
= 2\xi \int\!\!\! d^8z {\rm Tr}b {\overline b}    \ ,
\ena
where
$d{\tilde \mu} \equiv {\cal D}V{\cal D}c{\cal D}{\overline c}'{\cal D}{\overline c}{\cal D}c'{\cal D}b{\cal D}{\overline b}$. Obviously, (\ref{eq:conventional-BRST-generating-functional-eq1}) and (\ref{eq:conventional-BRST-generating-functional-eq2}) are equivalent. In particular, 
in (\ref{eq:conventional-BRST-generating-functional-eq1}), the N-L field plays the role of the N-K ghost in (\ref{eq:conventional-BRST-generating-functional-eq2}). Although (\ref{eq:conventional-BRST-generating-functional-eq2}) is convenient, the $nilpotent$ BRST symmetry is manifest in (\ref{eq:conventional-BRST-generating-functional-eq1}).

In the conventional approach, the origin of the anomalies may be traced to the Jacobian of the path-integral measure.\cite{Fujikawa}
In order to derive the anomalous superconformal W-T identity from (\ref{eq:conventional-BRST-generating-functional-eq1}), we carry out a change of the integration variable as 
$V \rightarrow V + \delta^{(Q1)}_{\rm sc} V$. Then we obtain
\bea
\label{eq:conventional-W-T-eq1}
- \langle \delta^{(Q1)}_{\rm sc} S^{(2)}_{\rm tot} \rangle
+ ({\rm Jacobian}-1)   = 0 \ ,
\ena
where the expectation value $\langle ... \rangle$ is defined by the generating functional $Z$ in (\ref{eq:conventional-BRST-generating-functional-eq1}).  
The quantity $( {\rm Jacobian} - 1)$ for the change of the integration variable is evaluated as follows. We expand the vector superfield in a complete set 
$\{ \varphi^a_m (z) = \langle a, z | m \rangle \}$ as 
$ V(z) = \sum_{m} c_m \varphi^a_m (z)$ 
and define the path-integral measure ${\cal D}V = \prod_m dc_m$.
Then the Jacobian is given by
\bea
\label{eq:Jacobian-def-eq1}
{}
& {} & ( {\rm Jacobian} - 1 )          \nn
& {} & \qquad = 
\sum_m \int\!\!\! d^8z \varphi^a_m (z) \Bigl\{ \displaystyle{\frac{1}{4}}\Omega^\a ( {\overline {\cal D}}^2{\cal D}_\a )
+ \displaystyle{\frac{1}{4}}{\overline \Omega}_{\dot \a}( {\cal D}^2{\overline {\cal D}}^{\dot \a} ) \Bigr\}^{ab}_z \varphi_m^b (z)     \ , \nn
& {} & \qquad = 
\lim_{\Lambda \rightarrow \infty}  \int\!\!\! d^8z \Bigl\{ \displaystyle{\frac{1}{4}}\Omega^\a ( {\overline {\cal D}}^2{\cal D}_\a )
+ \displaystyle{\frac{1}{4}}{\overline \Omega}_{\dot \a}( {\cal D}^2{\overline {\cal D}}^{\dot \a} ) \Bigr\}^{ab}_z
\langle b, z| e^{\square_\xi / \Lambda^2 } |a, z' \rangle |_{z=z'}       \ .
\ena
Here, the Gaussian regularization is introduced through the substitution 
$
\varphi^a_m (z) \varphi^a_m (z) 
\rightarrow 
\varphi^a_m (z) e^{- \lambda_m/\Lambda^2 } \varphi^a_m (z) 
$, 
with the eigenvalue $\lambda_m$, which satisfies the eigenvalue equation 
$
( - \square_\xi )^{ab} \varphi^b_m (z) = \lambda_m \varphi^a_m (z) 
$.
As is clear from (\ref{eq:anomaly-SQM-eq1}), the Jacobian is given by the same expression as in (\ref{eq:anomaly-SQM-eq1}). 
The expression for the Jacobian factor is also identical to the anomaly term in (\ref{eq:conventional-BRSTinvariant-W-T-eq2}), which is derived from a S-D equation. Namely, 
\bea
\label{eq:contact-term-def-eq1}
& {} &
 \Big\langle \int\!\!\! d^8z 
\displaystyle{\frac{\delta}{\delta V(z)}}  
 \delta_{\rm sc} V(z) 
 \Big\rangle          \nn
& {} & \qquad = 
\sum_m \int\!\!\! d^8z \varphi^a_m (z) \Bigl\{ \displaystyle{\frac{1}{4}}\Omega^\a ( {\overline {\cal D}}^2{\cal D}_\a )
+ \displaystyle{\frac{1}{4}}{\overline \Omega}_{\dot \a}( {\cal D}^2{\overline {\cal D}}^{\dot \a} ) \Bigr\}^{ab}_z \varphi_m^b (z)     \ .
\ena
Therefore, we have proved (\ref{eq:stochastic-BRST-correspondence-eq1}) in the one-loop approximation. 

From the
effective equation of motion,
$\delta^{(B)}_{\rm sc}\Gamma
= \delta^{(B)}_{\rm sc}S^{(0)} + \langle \delta^{(B)}_{\rm sc} S^{(2)}_{\rm tot} \rangle = 0$, which means
$
\langle \delta^{(Q2)}_{\rm sc}S^{(1)} \rangle = 0
$
in (\ref{eq:conservation-R2-eq1}),
 (\ref{eq:conventional-W-T-eq1}) reads
\bea
\label{eq:conventional-W-T-eq2}
& {} &
  \displaystyle{\frac{1}{2}} \int\!\!\! d^8z
( D^\a {\overline \Omega}^{\dot \a} - {\overline D}^{\dot \a} \Omega^\a ) \Big( R^{(0)}_{\a{\dot \a}} + \langle R^{(2)}_{\a {\dot \a}} \rangle  \Big)   
+ ({\rm Jacobian} -1)         \nn
& {} & \quad
= - i \!\!\int\!\!\! d^8z \Big\langle
\delta_{\rm BRST} \Big\{
( \delta_{\rm sc}^{(B)} + \delta_{\rm sc}^{(Q1)} ) {\rm Tr}((c' + {\overline c}' ) V) \Big\} \Big\rangle
+ \langle \delta^{(B)}_{\rm sc} S_{\rm NL} \rangle    \ .
\ena
Here we have used the relation 
$[\delta^{(B)}_{\rm sc} + \delta^{(Q)}_{\rm sc}, \ \delta_{\rm BRST} ] = 0 $.
We have also assumed that the superconformal transformations of the auxiliary fields and ghosts are induced by the background field dependence as
\bea
\label{eq:induced-background-transf}
\delta^{(B)}_{\rm sc} \varphi
& = &
[ ( \delta^{(B)}_{\rm sc} e^{g{\bf \Omega}^\dagger} )
e^{-g{\bf \Omega}^\dagger} , \    \varphi  ]        \ ,   \nn
\delta^{(B)}_{\rm sc} {\bar \varphi}
& = &
[ ( \delta^{(B)}_{\rm sc} e^{-g{\bf \Omega}} ) e^{g{\bf \Omega}} , \
  {\bar \varphi} ]       \  , 
\ena
for 
$\varphi = ( c, c', B, b )$
and
${\bar \varphi} = ({\bar c}', {\bar c}, {\bar B}, {\bar b} )$. 
In (\ref{eq:conventional-W-T-eq2}),
the terms that include the gauge fixing and ghosts are BRST exact and vanish in the (BRST invariant) vacuum expectation value. $\langle \delta^{(B)}_{\rm sc} S_{\rm NL} \rangle = 0$ holds, because $\langle B(z){\overline B}(z')\rangle = 0$. Thus, (\ref{eq:conventional-W-T-eq2}) yields the anomalous superconformal W-T identity (\ref{eq:SYM-anomalous-W-T-eq1}). This is a standard procedure to derive the superconformal anomaly. We have also confirmed that the S-D equation approach yields the anomalous superconformal W-T identity (\ref{eq:conventional-BRSTinvariant-W-T-eq2}) in the one-loop approximation. 

Next, we clarify the correspondence between the stochastic W-T identity (\ref{eq:BFM-stochastic-W-T-eq1}) with the Gaussian cut-off regularization and the conventional one (\ref{eq:conventional-W-T-eq2}). For this purpose, we specify the regularization procedure in the conventional approach. The regularization was introduced in (\ref{eq:Jacobian-def-eq1}) to evaluate the Jacobian factor of the path-integral measure. This prescription naturally introduces a Gaussian cut-off regularization for the vector superpropagator: 
\bea
\label{eq:vector-superpropagator-reg-eq1} 
\langle V^a ( z ) V^b ( z') \rangle_{\rm reg}     
= 
- \displaystyle{\frac{1}{2}} \langle a,z | (- {\hat \square}_\xi)^{-1} e^{{\hat \square}_\xi / \Lambda^2}   | b,z' \rangle        \ . \nn 
\ena
Furthermore, we are able to specify the regularization procedure of the superpropagators for the auxiliary field and ghost fields by using the Schwinger-Dyson and Slavnov-Taylor identities. 
From the S-D equation 
$
\int\!\!d\mu {\delta\over \delta B(z)}( V(z') e^{- S_{\rm tot}} ) = 0
$,
we $require$ the relation 
\bea
\label{eq:SQM-BFM-S-D-eq1}
\Big\langle B^a (z) V^b (z') \Big\rangle_{\rm reg}
= 2\xi \left( \displaystyle{\frac{1}{16}}{\overline {\cal D}}^2{\cal D}^2 \right)^{ac}_z
\Big\langle
V^c (z) V^b (z')
\Big\rangle_{\rm reg}    \ .
\ena
Similarly, $\langle {\overline B}^a (z) V^b (z') \rangle_{\rm reg}$ must satisfy the relation 
\bea
\label{eq:SQM-BFM-S-D-eq2}
\Big\langle {\overline B}^a (z) V^b (z') \Big\rangle_{\rm reg}
= 2\xi ( \displaystyle{\frac{1}{16}}{\cal D}^2{\overline {\cal D}}^2 )^{ac}_z
\Big\langle
V^c (z) V^b (z')
\Big\rangle_{\rm reg}    \ .
\ena
 These relations and the Slavnov-Taylor identities of the BRST symmetry enable us to express the background dependent tree level ghost superpropagators in terms of the regularized vector superpropagator. For example, we require that the S-T identity
$
\langle \delta_{\rm BRST}(c'(z)V(z')) \rangle = 0
$ 
holds under the regularization. This requirement and  (\ref{eq:SQM-BFM-S-D-eq1}) yield
\bea
\label{eq:ghost-vector-identity-eq1}
\Big\langle
c^{' a}(z) {\overline c}^b (z')
\Big\rangle_{\rm reg} 
& = & 
\Big\langle
c^a(z) {\overline c}^{' b} (z')
\Big\rangle_{\rm reg}             
 =  
2 \Big\langle B^a (z) V^b (z') \Big\rangle_{\rm reg}  \ , \nn    
&= & 
-2 \langle a,z | (\displaystyle{\frac{1}{4}}{\hat {\overline \nabla}}^2)(-{\hat \square}^{\rm FP})^{-1} e^{\xi {\hat \square}^{\rm FP}/ \Lambda^2} (\displaystyle{\frac{1}{4}}{\hat \nabla}^2)| b, z' \rangle      \,  \nn  
& = &
4\xi \Big\langle
\Big( \displaystyle{\frac{1}{4}} {\overline {\cal D}}^2_z V(z)\Big)^a
\Big( \displaystyle{\frac{1}{4}} {\cal D}^2_{z'} V(z') \Big)^b
\Big\rangle_{\rm reg}     \ .
\ena   
$
\langle
{\overline c}'(z) c(z')
\rangle_{\rm reg}
$
 and
$
\langle
{\overline b}(z) b(z')
\rangle_{\rm reg}
$
  are also expressed in terms of the regularized vector superpropagator in a similar manner: 
\bea
\label{eq:ghost-vector-identity-eq2}
\Big\langle
{\overline c}^{' a}(z) c^b (z')
\Big\rangle_{\rm reg} 
& = &
\Big\langle
{\overline c}^a(z) c^{' b} (z')
\Big\rangle_{\rm reg}               \  \nn    
&= & 
2 \langle a,z | \left( \displaystyle{\frac{1}{4}}{\hat \nabla}^2 \right)(-{\hat {\tilde \square}}^{\rm FP})^{-1} e^{\xi {\hat {\tilde \square}}^{\rm FP}/ \Lambda^2} \left( \displaystyle{\frac{1}{4}}{\hat {\overline \nabla}}^2 \right)| b, z' \rangle      \,  \nn  
& = &
- 4\xi \Big\langle
\Big( \displaystyle{\frac{1}{4}} {\cal D}^2_z V(z)\Big)^a
\Big( \displaystyle{\frac{1}{4}} {\overline {\cal D}}^2_{z'} V(z') \Big)^b
\Big\rangle_{\rm reg}     \     ,      \nn 
\Big\langle
b^a(z) {\bar b}^b (z')
\Big\rangle^\sim_{\rm reg}        
& = & 
 \Big\langle
\Big( \displaystyle{\frac{1}{4}} {\overline {\cal D}}^2_z V(z)\Big)^a
\Big( \displaystyle{\frac{1}{4}} {\cal D}^2_{z'} V(z') \Big)^b
\Big\rangle_{\rm reg}     \     .
\ena     
Here, the expectation value $\langle  ... \rangle^\sim$ of the N-K ghost superpropagator is defined by the generating functional ${\tilde Z}$ in (\ref{eq:conventional-BRST-generating-functional-eq2}). 
It should be noted that the regulator of the F-P ghost superpropagator must be $\Lambda^2/\xi$ for an arbitrary gauge parameter $\xi$. This is a consequence of the S-D equations and the S-T identities. The regularization prescription we have explained here for the conventional BRST invariant approach is presented in Ref.\citen{HOOS}. 
With these relations, in the regularized expectation value, we are able to make the substitutions 
\bea
\label{eq:substitution-prescription-eq1}
B^a 
&\rightarrow& 
\displaystyle{\frac{\xi}{8}}{\overline {\cal D}}^2 {\cal D}^2 V ^a    \ , \nn
{\overline B}^a 
&\rightarrow& 
\displaystyle{\frac{\xi}{8}}{\cal D}^2 {\overline {\cal D}}^2 V ^a    \ , \nn
\varphi^a 
&\rightarrow& 
- \displaystyle{\frac{1}{4}}{\overline {\cal D}}^2 V ^a    \ , \nn
{\overline \varphi}^a 
&\rightarrow& 
- \displaystyle{\frac{1}{4}}{\cal D}^2 V ^a    \ , 
\ena
where 
$\varphi \equiv ( c, c', b)$ and 
${\overline \varphi} \equiv ( {\bar c}', {\bar c}, {\bar b} )$. 
In this prescription for the substitutions, we assign each factor, $4\xi$, $4\xi$ and $1$ for each combination, $c'{\bar c}$, $c{\bar c}'$ and $b{\bar b}$, respectively. 
 
The BRST exact term in (\ref{eq:conventional-W-T-eq2}) consists of two parts. One is $\delta^{(B)}_{\rm sc}( S_{\rm gf} + S_{\rm FP})$ and the other is $\delta^{(Q)}_{\rm sc}( S_{\rm gf} + S_{\rm FP})$.
From (\ref{eq:SQM-BFM-S-D-eq1}) and (\ref{eq:ghost-vector-identity-eq1}), we find the correspondence between these terms and those in (\ref{eq:BFM-stochastic-W-T-eq1}): 
\bea
\label{eq:correspondence-gauge-fixing-eq1}
& {} &
 \int\!\!\! d^8z \Big\{
{\overline \Omega}^{\dot \a}  \Big\langle {\rm Tr} {\overline W}_{\dot \a}^{(0)} [V,\ \phi ] \Big\rangle^\infty_{\rm reg}   -  \Omega^\a \Big\langle {\rm Tr} [V,\ {\overline \phi} ] W_\a^{(0)}
    \Big\rangle^\infty_{\rm reg}     \Big\}   \nn
& {} & \qquad \qquad \qquad     
= 
i \int\!\!\! d^8z
\Big\langle
\delta^{(B)}_{\rm sc} {\rm Tr}\Big( ({\overline B} + B)V + \displaystyle{\frac{1}{2}}( {\overline c}c' + {\overline c}'c ) \Big)
\Big\rangle_{\rm reg}                    \ , \nn
& {} &
\displaystyle{\frac{1}{8}}\int\!\!\! d^8z \Big\{
{\overline \Omega}^{\dot \a}  \Big\langle {\rm Tr} \Big(
 ({\cal D}^2 {\overline {\cal D}}_{\dot \a} V )( {\overline \phi} - \phi )   \Big) \Big\rangle^\infty_{\rm reg}  
 - \Omega^\a \Big\langle {\rm Tr} \Big( ( {\overline \phi} - \phi )({\overline {\cal D}}^2 {\cal D}_\a V )   \Big) \Big\rangle^\infty_{\rm reg}   \Big\}         \nn
& {} & \qquad \qquad \qquad
=  
i \int\!\!\! d^8z \Big\langle \delta^{(Q1)}_{\rm sc} {\rm Tr}\Big( ({\overline B} + B) V\Big) \Big\rangle_{\rm reg}         \  . 
\ena
These relations hold without use of the classical background equations of motion. 
From (\ref{eq:correspondence-gauge-fixing-eq1}), we reinterpret the identity (\ref{eq:identity-vector-superpropagator-eq1}) in the stochastic superconformal W-T identity (\ref{eq:BFM-stochastic-W-T-eq1}) as
\bea
\label{eq:SQM-BRST-correspondence-eq1}
& {} &
[ {\rm the\ third\ and\ fourth\ terms\ on\ the\ r.h.s.\ of}\
(\ref{eq:identity-vector-superpropagator-eq1})  ]     \  \nn
& {} & \qquad = 
i \beta \Delta\t \int\!\!\! d^8z \Big\langle
\delta_{\rm BRST} \Big\{
( \delta_{\rm sc}^{(B)} + \delta_{\rm sc}^{(Q1)} ) {\rm Tr}((c' + {\overline c}' ) V) \Big\} \Big\rangle_{\rm reg}         \  . 
\ena
This is a proof of the second identity (\ref{eq:stochastic-BRST-correspondence-eq2}) in the one-loop approximation. 
Therefore, in the analysis of the anomalous superconformal W-T identity, (\ref{eq:stochastic-W-T-eq1}) and (\ref{eq:BFM-stochastic-W-T-eq1}), the mechanism by which the contribution from the stochastic gauge fixing term vanishes in the SQM approach is equivalent to the BRST exactness of the contributions from the gauge fixing and ghosts in the conventional approach. In particular, as is clear from (\ref{eq:identity-vector-superpropagator-eq1}), it is not necessary to fine tune the regulator of the F-P ghost superpropagator in the SQM approach. 

We finally show the equivalence of the effective stochastic equation of motion (\ref{eq:SQM-effective-eq-motion-eq1}) and the conventional one. 
It is more transparent to compare (\ref{eq:SQM-effective-eq-motion-eq1}) with the effective equations of motion derived from the familiar form of the generating functional  ${\tilde Z}$ in (\ref{eq:conventional-BRST-generating-functional-eq2}). 
With the help of the substitutions appearing in (\ref{eq:substitution-prescription-eq1}), it is easy to show that the background variation of the gauge fixing and ghost ( F-P and N-K ) terms is expressed as 
\bea
\Big\langle \delta^{(B)} ( {\tilde S}_{\rm gf} + S^{(2)}_{\rm FP} + S_{\rm NK})    
\Big\rangle^\sim_{\rm reg}  
& = & 
- \displaystyle{\frac{\xi}{8}} \Big\langle \int\!\!\! d^8z {\rm Tr}\Big\{
e^{-g{\bf \Omega}} ( \delta^{(B)} e^{g{\bf \Omega}} )
 [V,\  {\overline {\cal D}}^2{\cal D}^2  V  ]  \nn 
& {} & \qquad 
- ( \delta^{(B)} e^{g{\bf \Omega}^\dagger} ) e^{-g{\bf \Omega}^\dagger}   
 [V, \ {\cal D}^2{\overline {\cal D}}^2  V ]     
\Big\}    \Big\rangle_{\rm reg}         ,  
\ena
where the variations of the auxiliary fields and ghosts with respect to the background vector superfield are induced by a relation of the same form as (\ref{eq:induced-background-transf}). 
Then, (\ref{eq:SQM-effective-eq-motion-eq1}) reads
\bea
\label{eq:SQM-effective-eq-motion-eq2} 
{} 
& {} & 
[ {\rm the\ second\ term\ on\ the\ l.h.s\ of\ (\ref{eq:SQM-effective-eq-motion-eq1})} ]                     \nn 
& {} & \qquad =       
\Big\langle \delta^{(B)} ( S^{(2)} + {\tilde S}_{\rm gf} + S^{(2)}_{\rm FP} + S_{\rm NK})    
\Big\rangle^\sim_{\rm reg}        \  . 
\ena
This relation (\ref{eq:SQM-effective-eq-motion-eq2}) is, of course, equivalent to 
$\langle \delta^{(B)} ( S^{(2)} + S_{\rm gf} + S^{(2)}_{\rm FP} + S_{\rm NL}) \rangle_{\rm reg}$, 
derived from the generating functional $Z$ in (\ref{eq:conventional-BRST-generating-functional-eq1}).
In the conventional BFM, the generating functional of the 1-P-I vertices as the functional of the background fields is equivalent to the standard one with an unusual gauge fixing condition.\cite{Abbott} Therefore, the effective equation of motion for the background,
$\delta^{(B)}\Gamma = \delta^{(B)} S^{(0)} + \langle \delta^{(B)} {\tilde S}_{\rm tot}^{(2)} \rangle = 0$, is equivalent to the standard one. 
In order to derive the identity (\ref{eq:SQM-effective-eq-motion-eq2}), we have assumed that the background field is independent of the stochastic time and imposed the classical background equations of motion at finite stochastic time. Even without this requirement, we would have at most some discrepancies in (\ref{eq:SQM-effective-eq-motion-eq2}) that depend on ${\dot {\bf \Omega}}$ and ${\dot {\bf \Omega}}^\dagger$. Therefore, it is safe to conclude that the stochastic-time independence of the background fields,    
${\dot {\bf \Omega}} = {\dot {\bf \Omega}}^\dagger = 0$, 
is the necessary and sufficient condition to recover the standard effective equations of motion in equilibrium. 
The relation (\ref{eq:SQM-effective-eq-motion-eq2}) also shows that, at least in the one-loop approximation of the stochastic BFM, the stochastic gauge fixing procedure is equivalent to the Faddeev-Popov prescription in the conventional BFM which needs the N-K ghost as well as the F-P ghosts. This is consistent with the formal proof of this equivalence and the perturbative analysis in terms of the stochastic action principle.\cite{Nakazawa3}

\section{Discussion}

In this note, we have derived the anomalous W-T identity for the superconformal symmetry in SYM$_4$ in the context of the SQM approach. The superconformal anomaly comes from the contact term proportional to 
$\langle \Delta V \Delta V \rangle^\t_{\rm reg}$, 
while there is no contribution from the stochastic gauge fixing term. 
The self-cancellation mechanism of the contribution from the stochastic gauge fixing term corresponds to the BRST exactness that is exhibited by the conventional BRST invariant approach. 
Although our analysis is restricted in the one-loop approximation, it is a consequence of the fact that the stochastic gauge fixing procedure is introduced through the generator of the local gauge transformation into the time evolution equation of dynamical variables. Therefore, the expectation values of the local gauge invariant observables are independent of the stochastic gauge fixing procedure.

We have also studied the effective equation of motion in the context of the SQM approach. The S-D equation, i.e. the expectation value of the Langevin equation in BFM, recovers the standard effective equation of motion defined by the 1-P-I vertices. These analysis of SYM$_4$ in the superfield formalism explicitly confirms the equivalence of the SQM approach and the conventional BRST invariant one.

\section*{Acknowledgements}

This work was completed at KEK. The author would like to thank all the members of the theory group at KEK for their hospitality. He wishes to thank H. Suzuki for fruitful discussions. He also wishes to thank H. Kawai and T. Tada for encouragement.


\end{document}